\documentclass[journal=aamick,manuscript=article]{achemso}

\usepackage{chemformula} % Formula subscripts using \ch{}
\usepackage[T1]{fontenc} % Use modern font encodings
\usepackage{graphicx}
\usepackage{fancyhdr}
\usepackage{makeidx}
\usepackage{amssymb,amsmath}
\usepackage{physics}
\usepackage{upquote}
\usepackage{pdfpages}
\usepackage{xcolor}
\usepackage{enumerate}
\newcommand{\bd}{\begin{document}}
\newcommand{\ed}{\end{document}}
\newcommand{\bc}{\begin{center}}
\newcommand{\ec}{\end{center}}
\newcommand{\vs}{\vspace}
\newcommand{\hs}{\hspace}
\newcommand{\beq}{\begin{equation}}
\newcommand{\eeq}{\end{equation}}
\newcommand{\beqs}{\begin{eqn*}}
\newcommand{\eeqs}{\end{eqn*}}
\newcommand{\bq}{\begin{quote}}
\newcommand{\eq}{\end{quote}}
\newcommand{\lb}{\linebreak}
\newcommand{\mb}{\makebox}
\newcommand{\fb}{\framebox}
\newcommand{\mc}{\multicolumn}
\newcommand{\ben}{\begin{enumerate}}
\newcommand{\een}{\end{enumerate}}
\newcommand{\bit}{\begin{itemize}}
\newcommand{\eit}{\end{itemize}}
\newcommand{\ov}{\overline}
\newcommand{\un}{\underline}
\newcommand{\lt}{\left}
\newcommand{\rt}{\right}
\newcommand{\ba}{\begin{array}}
\newcommand{\ea}{\end{array}}
\newcommand{\beqa}{\begin{eqnarray}}
\newcommand{\eeqa}{\end{eqnarray}}
\newcommand{\beqas}{\begin{eqnarray*}}
\newcommand{\eeqas}{\end{eqnarray*}}
\newcommand{\bfg}{\begin{figure}}
\newcommand{\efg}{\end{figure}}
\newcommand{\pad}{\partial}
\newcommand{\nn}{\nonumber}
\newcommand{\la}{\leftarrow}
\newcommand{\ra}{\rightarrow}
\newcommand{\lgla}{\longleftarrow}
\newcommand{\lgra}{\longrightarrow}
\newcommand{\La}{\Leftarrow}
\newcommand{\Ra}{\Rightarrow}
\newcommand{\Lra}{\Leftrightarrow}
\newcommand{\Lgla}{\Longleftarrow}
\newcommand{\Lgra}{\Longrightarrow}
\renewcommand{\a}{\alpha}
\renewcommand{\b}{\beta}
\newcommand{\g}{\gamma}
\newcommand{\G}{\Gamma}
\renewcommand{\d}{\delta}
\newcommand{\D}{\Delta}
\newcommand{\e}{\epsilon}
\newcommand{\eps}{\epsilon}
\newcommand{\s}{\sigma}
\renewcommand{\l}{\lamda}
\newcommand{\m}{\mu}
\newcommand{\n}{\nu}
\renewcommand{\S}{\Sigma}
\newcommand{\p}{\pi}
\newcommand{\om}{\omega}
\newcommand{\Om}{\Omega}
\newcommand{\tri}{\triangle}
\newcommand{\ti}{\times}
\newcommand{\f}{\frac}
\newcommand{\ds}{\displaystyle}
\newcommand{\bm}[1]{\mb{{\boldmath $#1$}}}
\newcommand{\alter}[2]{\lt\{ \ba{ll}#1 \\ #2 \ea \rt.}
\newcommand{\alt}[4]{\lt\{ \ba{ll}#1 & \mb{if \, \,}#2 \\ #3 & \mb{}#4 \ea
    \rt.}
\newcommand{\altn}[4]{\lt\{ \ba{rl}#1 & \mb{if \, \,}#2 \\ #3 & \mb{}#4 \ea
    \rt.}
\newcommand{\altif}[4]{\lt\{ \ba{ll}#1 & \mb{if \, \,}#2 \\ #3 &
\mb{if \, \,}#4 \ea \rt.}
\newcommand{\altnif}[4]{\lt\{ \ba{rl}#1 & \mb{if \, \,}#2 \\ #3 &
\mb{if \, \,}#4 \ea \rt.}
\newcounter{algc}
\newcounter{romc}
\newcounter{Alphc}
\newcommand{\bl}{\begin{list}{{\it Step} ~\arabic{algc}~:} {\usecounter{algc}
                \setlength{\topsep}{0pt} \setlength{\itemsep}{0pt}}}
\newcommand{\el}{\end{list}}
\newcommand{\blr}{\begin{list}{~\roman{romc}~:} {\usecounter{romc}
                \setlength{\topsep}{0pt} \setlength{\itemsep}{0pt}}}
\newcommand{\elr}{\end{list}}
\newcommand{\bla}{\begin{list}{~\Alph{Alphc}~:} {\usecounter{Alphc}
                \setlength{\topsep}{0pt} \setlength{\itemsep}{0pt}}}
\newcommand{\ela}{\end{list}}
\newcommand{\tsup}{\textsuperscript}
\newcommand{\tsub}{\textsubscript}

\newtheorem{theorem}{Theorem}

\author{Krishna Murali}
\author{Nithin Abraham}
\author{Sarthak Das}
\author{Sangeeth Kallatt}
\author{Kausik Majumdar}
\email{kausikm@iisc.ac.in}
\affiliation{Department of Electrical Communication Engineering, \\Indian Institute of Science, Bangalore 560012, India}

\title{Highly Sensitive, Fast Graphene Photodetector with Responsivity $>10^6$ A/W Using Floating Quantum Well Gate}


\begin{document}

\begin{tocentry}

\begin{center}
\includegraphics[]{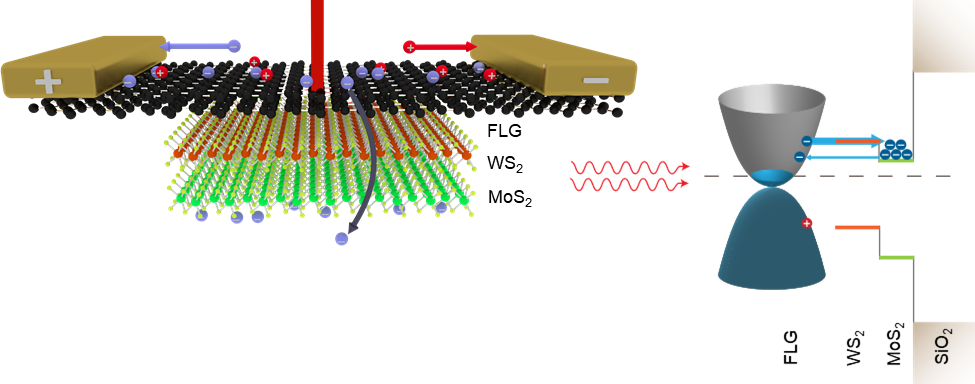}
\end{center}
\end{tocentry}

\begin{abstract}
  Graphene, owing to its zero bandgap electronic structure, is promising as an absorption material for ultra-wideband photodetection applications. However, graphene-absorption based detectors inherently suffer from poor responsivity due to weak absorption and fast photocarrier recombination, limiting their viability for low intensity light detection. Here we use a graphene/WS$_2$/MoS$_2$ vertical heterojunction to demonstrate a highly sensitive photodetector, where the graphene layer serves dual purpose, namely as the light absorption layer, and also as the carrier conduction channel, thus maintaining the broadband nature of the photodetector. A fraction of the photoelectrons in graphene encounter ultra-fast inter-layer transfer to a floating monolayer MoS$_2$ quantum well providing strong quantum confined photogating effect. The photodetector shows a responsivity of $4.4\times 10^6$ A/W at 30 fW incident power, outperforming photodetectors reported till date where graphene is used as light absorption material by several orders. In addition, the proposed photodetector exhibits an extremely low noise equivalent power ($N\!E\!P$) of $<4$ fW/$\sqrt{Hz}$ and a fast response ($\sim$ milliseconds) with zero reminiscent photocurrent. The findings are attractive towards the demonstration of graphene-based highly sensitive, fast, broadband photodetection technology.
\end{abstract}

\section{Introduction}
Photodetectors operating in the infra-red spectral regime find widespread applications across different areas, such as military, space, communication, and medical imaging, to name a few \cite{Tan2018,Martyniuk2014,Downs2013,Potter1962,Rogalski2011}. Technologies based on III-V semiconductors\cite{Craig2015,Gu2015}, Pb based quantum dots\cite{DeIacovo2016,Cui2006} and mercury cadmium telluride\cite{Hall2004,Lei2015} currently dominate this space. However, there is a notable research focused on simplifying the fabrication process, reducing material cost, using non-hazardous materials in terms of environment, health and safety, and improving the detector performance at room temperature, thereby eliminating additional cooling requirements.
\\
\\
Graphene can absorb light of extremely wide wavelength range (ultra-violet to terahertz wavelength regime) due to its unique gapless band structure\cite{Mak2012}. Graphene is thus widely considered as a promising material for broadband photodetection and hence is attractive for a plethora of applications \cite{Xia2009,Park2009,Wu2011,Furchi2012}. Further, the graphene-based photodetectors exhibit ultra-fast operation \cite{Xia2009} due to superior carrier mobility. Unfortunately, graphene photodetectors typically do not show high sensitivity due to two important issues. First, the ultra-thin nature of graphene impedes the overall light absorption efficiency. For example, a monolayer-thick graphene absorbs only about 2.3\% of the light intensity incident on it \cite{Nair2008}. Second, photo-excited carriers in graphene possess short lifetime which is on the order of picoseconds \cite{George2008,Rana2009}. One must separate the photo-generated electron-hole pairs within a time scale that is less than the photocarrier lifetime in order to generate a sizeable photocurrent. Consequently, a majority of the reported photodetectors in literature where graphene is used as light absorption medium, including graphene/metal junction \cite{Mueller2010,Yoo2018,Sassi2017}, p-n junction using graphene \cite{Wu2011,Freitag2013a}, and bilayer graphene under an electrical bias \cite{Spirito2014}, exhibit weak sensitivity.
\\
\\
In this context, there has been a considerable amount of effort towards achieving photodetectors with a large responsivity by using photogating effect  \cite{Chen2015a,Chang2018,Yu2013,Wan2017,Tao2017,Liu2014,Konstantatos2012,Roy2013,Yu2017,Lan2017,Wu2018,Xiong2019,Huo2017,Casalino2018}. However, these structures primarily use graphene as a photocarrier transport layer, while the light absorption occurs in a different material, for example, transition metal dichalcogenides \cite{Roy2013,Zhang2015c,Vabbina2015}, quantum dots \cite{Konstantatos2012,Sun2012,Pan2018} and perovskites \cite{Pan2018,Chang2017,Li2017a}. The corresponding threshold wavelength of the photo-absorbing material thus limits the operation range of the photodetector, and the broadband nature of the photodetection that graphene offers is completely lost. In addition this, these detectors employing photogating while can achieve a high-gain, suffer from slow response, and exhibit a large reminiscent photocurrent \cite{Konstantatos2012,Roy2013,Zhang2015c,Zhang2013,Tan2017} due to slow de-trapping of carriers. To mitigate this, it has been proposed in the past to flush the trapped carriers using a gate voltage pulse which can restore the initial state of the device \cite{Konstantatos2012,Roy2013}. However, such an arrangement requires additional circuits and thus increases the system level complexity. In order to eliminate these bottlenecks, in this work, we propose and demonstrate a novel photodetector that uses a vertical van der Waals (vdW) heterojunction comprising of graphene/WS$_2$/MoS$_2$ stack. In the proposed structure, graphene serves as the light absorbing layer and the carrier transport layer, while the MoS$_2$ quantum well serves as a floating gate. The device simultaneously achieves ultra-high responsivity, extremely low noise, fast photodetection with no reminiscent photocurrent at room temperature over a broad wavelength range facilitating broadband operation.
\\
\\

\section{Results and discussions}

\subsection{Principle of operation}

The architecture and working principle of the proposed device is explained schematically in Fig. 1a-c. The vertical stack consists of a monolayer (1L) or bilayer (2L) WS$_2$ film sandwiched between few-layer graphene (FLG) on top and 1L-MoS$_2$ at the bottom. The whole stack is placed on Si substrate coated with 285 nm thermally grown SiO\tsub2. MoS$_2$ having a conduction band offset with WS$_2$, as shown in Fig. 1c, creates an ultra-thin quantum well for electrons, with SiO\tsub2 on the other side. Under light illumination with photon energy less than the excitonic gap of MoS\tsub2 (and WS\tsub2), electron-hole pairs are generated only in the graphene layer. Since inter-layer carrier transfer process in vdW heterojunction is extremely fast ($\sim$ sub-ps) \cite{Massicotte2016,Hong2014,Chen2016,Hill2017}, a fraction of the photo-generated electrons are transferred to the MoS$_2$ quantum well through the ultra-thin WS$_2$ layer before recombination in graphene (Fig. 1b and left panel of Fig. 1c). Depending on the wavelength of excitation, both tunneling and over-the-WS$_2$-barrier transfer processes are possible. The electrons, after being transferred to MoS$_2$, swiftly thermalize to the conduction band edge of MoS$_2$. A fraction of these carriers can also be captured by the sub-bandgap trap states \cite{Nan2014} in MoS$_2$. The trapped carriers in the MoS$_2$ quantum well can in further tunnel through the WS$_2$ barrier layer to come back to graphene. These population and depopulation mechanisms are quite fast due to the ultra-fast nature of the inter-layer charge transfer, which help to quickly create a steady state photoelectron population in the MoS$_2$ floating quantum well. The WS$_2$ spacer being an ultra-thin layer allows strong electrostatic coupling between the MoS$_2$ layer and the graphene channel. The MoS$_2$ quantum well with the captured photoelectrons thus acts as a floating gate on the graphene channel, modulating the Dirac point and the in-plane conductivity of the FLG film, in turn generating a measurable change in the current as detected by the electrodes shown in Figs. 1a-b.
\\
\\
Note that there is a strong built-in field due to the offset of the conduction band edges between MoS$_2$ and graphene, as shown in the band diagram in Fig. 1c. Such field favors a fast inter-layer transfer of the stored photoelectrons in MoS$_2$ to graphene. When the light source is turned off (Fig. 1c, right panel), the captured photoelectrons in the MoS$_2$ quantum well thus swiftly tunnel through the WS$_2$ barrier layer to graphene, and thus switching off the gating action immediately. This mechanism allows us to completely suppress the reminiscent photocurrent. Further, we do not need an external gate pulse to flush the confined photoelectrons. The detector thus is expected to exhibit a fast response due to the intrinsically small time scale of the inter-layer charge transfer and the high in-plane carrier mobility in graphene.
\\
\\

\subsection{Evidence of carrier storage in MoS$_2$ floating quantum well}

Fig. 2a shows the optical image of a fabricated FLG/1L-WS$_2$/1L-MoS$_2$ stack, along with isolated 1L-WS$_2$ and 1L-MoS$_2$ portions as controls. The Raman shifts from the isolated portions, as well as from the junction are shown in Fig. 2b. Under illumination, to demonstrate the charge storage in the bottom monolayer MoS$_2$ without immediately being transferred to graphene through WS$_2$, we employ photoluminescence (PL) quenching measurement. Figs. 2c-d show the PL intensity acquired from the isolated 1L-WS$_2$ (in orange), isolated 1L-MoS$_2$ (in green) and the junction (in black) separately at 150 K and 295 K with 532 nm laser excitation. At 150 K the monolayer WS$_2$ shows the exciton ($A_{1s}$) peak around 2.07 eV and the trion ($A_{1s}^T$) peak around 2.04 eV, whereas the monolayer MoS$_2$ exhibits the $A_{1s}$ peak around 1.94 eV, and the $A_{1s}^T$ peak at around 1.90 eV. On the junction, although there is no significant change in the peak position in the PL spectra (due to a combined effect of bandgap renormalization and exciton binding energy reduction \cite{Ugeda2014, Gupta2017}), the WS$_2$ PL shows almost 90\% to 95\% quenching of the exciton and the trion peaks. This indicates an ultra-fast inter-layer charge transfer from WS$_2$ to both graphene and MoS$_2$ before the radiative decay of the exciton. On the contrary, the MoS$_2$ peaks do not show any significant quenching in the PL spectra at the junction. At 295 K, the peaks are red shifted due to a reduction in bandgap, but the observations regarding quenching remain similar to that at 150 K.
\\
\\
Further, to avoid any possible compensation effect by photo-carrier injection from WS$_2$ to MoS$_2$, we reduce the temperature to 3.7 K, thus pushing up the excitonic gap of 1L-WS$_2$ to 2.1 eV, and selectively excite only the 1L-MoS$_2$ layer at the junction with 633 nm laser as the photon energy ($\sim$ 1.96 eV) is lower than the 1L-WS$_2$ optical bandgap. The results as shown in Fig. 2e suggest that PL intensity of the trion peak (the exciton peak being truncated by the edge filter due to resonant excitation) of MoS$_2$ is similar in isolated portion and at the junction. This unambiguously shows that it is possible to create a steady state population of stored photo-carriers in the MoS$_2$ quantum well without being immediately transferred to graphene - allowing for the proposed photogating action. To confirm the point further, we also perform circular polarization resolved PL experiment using near resonant excitation in MoS$_2$ by 633 nm laser at 3.7 K (Fig. 2f). We obtain a similar degree ($\sim$ 25\%, obtained by peak fitting with Voigt function) of circular polarization for the trion peak both on the isolated MoS$_2$ (left panel) and on the junction (right panel), which indicates that the valley polarization of MoS$_2$ trions remains intact at the junction. This further justifies suppression of additional scattering channel in MoS$_2$ at the junction owing to desirable carrier confinement. Note that, at low temperature, we observe strong sub-bandgap luminescence both from WS$_2$ and MoS$_2$, indicating presence of defect states \cite{Koperski2017,Plechinger2016,Nagler2018,Vaclavkova2018}, which play an important role in obtaining large photodetector gain, as explained later.
\\
\\

\subsection{Photodetector fabrication and characterization}

The photodetector (D1) fabrication process steps are illustrated in Fig. 3a. The device fabrication process starts with mechanical exfoliation of MoS$_2$ flakes over Si substrate coated with 285 nm SiO2. 1L-MoS$_2$ flakes are identified by optical contrast and Raman spectroscopy. The 2L-WS$_2$ flake is then transferred on top of the 1L-MoS$_2$ flake using a micromanipulator under optical microscope. For making MoS$_2$ island, we use electron beam lithography, followed by dry etch using chlorine chemistry. FLG flake is then transferred on the heterostructure. Electrodes for contacting the FLG film are then defined by a second electron beam lithography step, followed by the deposition of Ni (10nm)/ Au (50nm) electrodes by DC sputtering and subsequent lift-off. A third lithography step followed by a final etch step is used to form a regular pattern of the FLG film so that graphene falls only on top of the WS$_2$/MoS$_2$ junction area, avoiding any un-gated parallel conduction path. More details on the layer transfer and device fabrication are provided in the \textbf{Methods} section. The junction area of the fabricated device is $6$ $\mu$m$^2$.
\\
\\
To demonstrate the concept proposed, we choose few-layer graphene over monolayer as the active layer. While monolayer graphene exhibits more gate tunability over few-layer, the choice of few-layer is to (1) enhance light absorption, (2) suppress substrate trap induced noise, and (3) reduce ambience induced detrimental effects. We also choose 2L-WS$_2$ over 1L as the sandwiched barrier layer (1) to improve over-the-barrier injection from graphene to MoS$_2$ due to suppressed barrier height, and (2) to increase the tunnelling width for band edge electrons in MoS$_2$ to move back to graphene, thereby increasing the steady state population of captured photoelectron density in MoS$_2$ floating gate. Fig. 3b shows the PL spectra of 1L-MoS$_2$/2L-WS$_2$ stack at 295 K, and the band-offset between the two can be inferred from the different peaks, as illustrated in Fig. 3c. The free exciton peaks of MoS$_2$ (A and B) and WS$_2$ (A) are distinctly observed from the isolated portions. The 2L-WS$_2$ also clearly shows the indirect peak around 1.75 eV, indicating its bilayer nature. At the junction area, the exciton and indirect peaks of WS$_2$ are quenched whereas the MoS$_2$ peaks are distinctly discernable. We also observe a distinct inter-layer exciton peak at 1.45 eV, which is absent in the individual layers, indicating strong electronic coupling between the vdW layers.
\\
\\
The $I_d$-$V_g$ characteristics of the FLG under dark condition is shown in Fig. 4a, with a Dirac voltage ($V_D$) around $-10.8$ V indicating n-type doping. We operate the device as a photodetector keeping the external gate voltage as zero. The operating point under dark condition is indicated by point $P_1$. Under illumination, the captured photoelectrons by the MoS$_2$ quantum well electrostatically gates the FLG, pushing the device operating point towards left to $P_2$, providing a negative photocurrent, as schematically depicted in Fig. 1c. The top inset of Fig. 4a shows the drain voltage ($V_d$) dependent dark current under zero gate bias.
\\
\\
Fig. 4b shows the measured photocurrent ($I_{ph}=I_{light}-I_{dark}$) of the photodetector D1 as a function of $V_d$, for three different optical powers ($P_{op}$) incident on the junction, at 532 nm excitation wavelength. Figs. 4c-e show the transient response of the photodetector at 532 nm, 785 nm, and 851 nm, respectively, illustrating fast switching for the incident powers used. The measured rise and fall times are found to be $<$ 10 ms, and only limited by the resolution of the measurement equipment. We do not observe any reminiscent photocurrent in the transient characteristics, which are typically observed in high-gain detectors using photo-gating effect. As mentioned earlier, this is achieved by the inherent built-in field in the device in the vertical direction that swiftly pushes the captured photo-carriers to graphene channel when light is turned off. This eliminates the requirement for additional refreshing circuit to push the detector to its original state. Note that the noise level is higher for the 532 nm excitation, which limits the detection of low power optical signal, while the noise level for 785 nm and 851 nm excitation is much lower, even when the incident power is very low (bottom panels of Figs. 4d-e). 532 nm photons having energy higher than the excitonic bandgap of MoS$_2$ (and WS$_2$), can be directly absorbed by these layers. The photo-generated large hot carrier density in MoS$_2$ can in turn directly interact with the trap states in SiO2 substrate, causing a large noise. On the other hand, when excitation energy is smaller than MoS$_2$ excitonic gap (for example, 785 and 851 nm), only graphene absorbs the excitation, followed by photoelectron transfer to MoS$_2$ quantum well. During this process, the hot photo-carriers lose their energy, which significantly suppresses the interaction with the substrate traps states, reducing the gain noise and allowing us to detect optical signal down to 30 fW.
\\
\\
Using the above argument, the ultra-low noise level exhibited by the photodetector at sub-bandgap excitation of MoS$_2$ and WS$_2$ helps us to rule out a direct absorption by the defect states in these materials. Further, absorption of the near-infrared wavelengths used in this work would require involvement of deep level trap states of MoS$_2$ and WS$_2$. De-trapping of photocarriers from such deep trap states would be very slow, and would result in a slow decay of the photocurrent. The fast transient response observed from the detector thus corroborates that the effect of direct sub-bandgap absorption in MoS$_2$ and WS$_2$ layers can be ignored in our work.
\subsection{Performance evaluation, modeling, and benchmarking}

The extracted responsivity ($R=\frac{I_{ph}}{P_{op}}$) of the photodetector is plotted in Fig. 5a as a function of incident optical power density for different wavelengths of excitation at $V_d=0.5$ V. There are two remarkable observations from this plot. First, the magnitude of the responsivity, particularly at low power density, is extremely high, reaching $4.4\times10^6$ A/W at the minimum power density applied. This is several orders of magnitude higher than that of any reported detector based on light absorption by graphene, as shown in Table \ref{tb1}. Second, the responsivity follows the same trend line irrespective of the wavelength of the incident light, in agreement with broadband light absorption by graphene.
\\
\\
In order to get insights into such high gain mechanism in the detector, we construct a unified model that captures the density dependent recombination time of photo-carriers in graphene ($\tau_r$) \cite{Rana2009,Rana2007}, fast inter-layer transfer of photo-carriers from graphene to MoS$_2$ floating gate ($\tau_{GM}$) and vice versa ($\tau_{MG}$), and possible trapping ($\tau_{MT}$) and de-trapping times ($\tau_{TM}$) of the carriers \cite{Knobloch2018,Furchi2014} in the MoS$_2$ bandgap. The coupled differential equations that govern the photoinduced carrier densities in graphene ($\Delta n_G$) and MoS$_2$ ($\Delta n_M$), and the trapped carrier density in MoS$_2$ ($n_T$) are given by\\
\beq\label{eq:m1}
\frac{d\Delta n_G}{dt} = \eta\phi - \frac{\Delta n_G}{\tau_r(\Delta n_G)} - \frac{\Delta n_G}{\tau_{GM}} + \frac{\Delta n_M}{\tau_{MG}}
\eeq
\beq\label{eq:m2}
\frac{d\Delta n_M}{dt} = \frac{\Delta n_G}{\tau_{GM}} - \frac{\Delta n_M}{\tau_{MG}} + \frac{n_T}{\tau_{TM}} - \frac{\Delta n_M}{\tau_{MT}}\Big{(}1-\frac{n_T}{N_T}\Big{)}
\eeq
\beq\label{eq:m3}
\frac{dn_T}{dt} = \frac{\Delta n_M}{\tau_{MT}}\Big{(}1-\frac{n_T}{N_T}\Big{)} - \frac{n_T}{\tau_{TM}}
\eeq
where $N_T$ is the trap state density in MoS$_2$ per unit area, $\eta$ is internal quantum efficiency of few-layer graphene absorption, and $\phi$ is the photon flux. Under steady state condition, putting the left hand sides of equations (\ref{eq:m1})-(\ref{eq:m3}) to zero, and realizing that there is no net flow of charge from graphene to MoS$_2$, we analytically obtain $\Delta n_G = \eta\phi\tau_r(\Delta n_G)$, $\Delta n_M = \frac{\Delta n_G\tau_{MG}}{\tau_{GM}}$, and $n_T = \frac{\Delta n_M N_T}{\Delta n_M + N_T\frac{\tau_{MT}}{\tau_{TM}}}$. The responsivity can then be calculated as
\beq
R_{calculated} = \frac{q}{hc}\frac{\lambda g_m(\Delta n_M + n_T)}{AC_{ox}\phi}
\eeq
where $h$ is Plank constant, $q$ is absolute electron charge, $c$ is velocity of light, $\lambda$ is the excitation wavelength, $g_m$ is the extracted transconductance from Fig. 4a at operating point $P_1$, $A$ is device active area, and $C_{ox}$ is back gate oxide capacitance. The orange curve is the simulated responsivity with fitting parameters $\frac{\tau_{MG}}{\tau_{GM}}=10$, $\frac{\tau_{MT}}{\tau_{TM}}=5\times10^{-4}$, and $N_T=5\times10^{10}$ cm\tsup{-2}. In the same plot, we also show the calculated $R$ by turning off the trap states in the model, as denoted by the blue curve. At high incident optical power density, the traps in MoS$_2$ play a less important role and photocurrent saturation is dominated by reduction in $\tau_r$ in graphene with increasing photon flux. At low incident optical power density, as $\tau_{TM}>>\tau_{MG}$ and $\tau_{TM}>>\tau_{MT}$, the relatively slow de-trapping of carriers builds up larger electron density in the trap states than in the conduction band of the floating MoS$_2$ gate and thus traps play a major role in photogating and help obtaining the large responsivity. The simulated results follow the experimentally obtained $R$ remarkably well over 11 orders of magnitude of incident optical power density. The minor deviation of the simulated results from the data at smaller wavelength and high optical power density is because the model does not take into account direct absorption in MoS$_2$. The experimentally obtained $R$ is slightly lower from the model predicted trend line at longer wavelength and is likely due to reduced energy of hot electrons, which affects the injection efficiency from graphene to MoS$_2$.
\\
\\
In order to estimate the sensitivity of the device, we first extract the root mean squared noise current ($N_{R\!M\!S}$) integrated up to a detection bandwidth of 1 Hz, that is with an averaging time of 0.5 s. The electrical signal-to-noise ratio ($S\!N\!R$) is then obtained as
\beq
S\!N\!R = \frac{I_{ph}}{N_{R\!M\!S}}
\eeq
and plotted in Fig. 5c as a function of incident power for different wavelengths. The smallest power incident on the junction during the experiment is 30 fW (@851 nm), and even at such a low power, we obtain an $S\!N\!R$ of $\sim 10$. Both $532$ nm and $633$ nm excitations, which directly excite the MoS\tsub2 and WS\tsub2 layers at the junction, produce smaller $S\!N\!R$ compared to the larger wavelength excitations. This is attributed to an enhanced noise level due to stronger interaction with SiO\tsub2 traps. The noise equivalent power ($N\!E\!P$) is a figure of merit that is often used to characterize the sensitivity and hence the minimum detectable optical power incident on the photodetector. $N\!E\!P$ represents the minimum input power required to obtain unity $S\!N\!R$. The $N\!E\!P$ (normalized to 1 Hz) \cite{Mackowiak2007} of the detector is then calculated as
\beq\label{eq:NEP}
N\!E\!P = \frac{N_{R\!M\!S}}{R}|_{S\!N\!R = 1}
\eeq
Since we only reached $S\!N\!R=10$, from Equation \ref{eq:NEP}, we estimate that the $N\!E\!P$  of the photodetector is $< 4$ fW/$\sqrt{Hz}$. The detection limit, as defined by $S\!N\!R = 1$ is shown in Fig. 5c by the dashed line. We have shown the performance of another fabricated FLG/2L-WS$_2$/1L-MoS$_2$ heterojunction photodetector (D2) in \textbf{Supporting Information S1}. D2 exhibits a maximum responsivity of $2.29 \times 10^6$ A/W, and $N\!E\!P < 10.9$ fW/$\sqrt{Hz}$ at an excitation wavelength of 851 nm.
\\
\\
We benchmark the performance of the photodetector with existing photodetectors using graphene as light absorption medium. We note from the table that the detectors exhibiting high speed of operation suffers from low responsivity, while devices with larger gain show slow response. In this regard, the detector reported in this work not only outperforms any existing graphene-absorption based detector in terms of responsivity by several orders of magnitude, it also simultaneously achieves an extremely low $N\!E\!P$ and a fast response.
\begin{table}
\centering
\caption{ BENCHMARKING TABLE}
\label{tb1}
\resizebox{\textwidth}{!}{\begin{tabular}{|c|c|c|c|c|}
 \hline
  \textbf{Device description} & \textbf{Absorbing medium} & \textbf{R (A/W)} & \textbf{NEP (W/$\sqrt{Hz}$)} & \textbf{Response time (s)} \\
  \hline
  \textbf{FLG/WS\tsub{2}/MoS\tsub{2} (This work)} & \textbf{Graphene} & $\mathbf{4.4 \times 10^6}$ & $\mathbf{< 4 \times 10^{-15}}$ & $\mathbf{< 10^{-2}}$ \\
  Metal/G/Metal detector \cite{Mueller2010} & Graphene & $6.11 \times 10^{-3}$ & -& $25 \times 10^{-12}$ \\
  Suspended graphene \cite{Freitag2013a} & Graphene & $10 \times 10^{-3}$ & - & - \\
  Graphene with THz antenna \cite{Spirito2014} & Graphene & $1.3 \times 10^{-3}$ & $2 \times 10^{-9}$ & - \\
  G/Ta\tsub{2}O\tsub{5}/G \cite{Liu2014} & Graphene & $\sim 10^3$ & $10^{-11}$ & $\sim 1$ \\
  G/MoS\tsub{2} \cite{Vabbina2015} & Graphene & $1.26$ & $7.8 \times 10^{-12}$ & - \\
  Metal/G/Metal with GQD arrays \cite{Zhang2013} & Graphene & $8.61$ & - & $\sim 100$ \\
  Plasmonically enhanced graphene \cite{Ma2018}  & Graphene & $0.5$ & - & $9 \times 10^{-12}$ \\
  Biased Graphene \cite{Freitag2013}& Graphene & $2.5 \times 10^{-4}$ & - & - \\
  \hline
\end{tabular}}
\end{table}
\section{Conclusion}
In summary, we have proposed and experimentally demonstrated a new class of highly sensitive photodetector that utilizes graphene as both light absorption medium as well as conduction channel and exploits ultra-fast inter-layer charge transfer to a floating quantum well enabling strong photogating. Graphene being the light absorbing medium, the photodetection principle is appealing for operation over a large wavelength range. The uniqueness of the device is demonstrated by eliminating the weak sensitivity observed in typical graphene-absorption based detectors. The proposed device achieves significantly superior performance over existing graphene-absorption-based photodetector reports in terms of simultaneously maintaining ultra-high responsivity, high signal to noise ratio at ultra-low incident power, and fast transient response, while reducing system level complexity by automatic flushing of captured photo-carriers when illumination is turned off. The photodetector also holds a key advantage in terms of maintaining room temperature operation, and using low-cost, less hazardous materials. The findings are appealing for highly sensitive broadband photodetection applications.

\section{Methods}

\textbf{Device fabrication:} The photodetector fabrication process steps are illustrated in Fig. 3a. The device fabrication process starts with mechanical exfoliation of MoS$_2$ flakes over Si substrate coated with 285 nm SiO2. 1L-MoS$_2$ flakes are identified by optical contrast and Raman spectroscopy. We then exfoliate WS$_2$ flakes on PDMS sheet and identify 2L-WS$_2$ using optical contrast and confirm with Raman and PL signal. The 2L-WS$_2$ flake is then transferred on top of the 1L-MoS$_2$ flake using a micromanipulator under optical microscope. For better adhesion, we heat this stack at 75$^\circ$C for 2 minutes on a hot plate. For making MoS$_2$ island, we use electron beam lithography, followed by dry etch using chlorine chemistry. FLG flake is then transferred on the heterostructure using a similar technique discussed above, followed by another heating step. Electrodes for contacting the FLG film are then defined by a second electron beam lithography step, followed by the deposition of Ni (10nm)/ Au (50nm) electrodes by DC sputtering and subsequent lift-off. A third lithography step followed by a final etch step is used to form a regular pattern of the FLG film so that graphene falls only on top of the WS$_2$/MoS$_2$ junction area, avoiding any un-gated parallel conduction path.

\begin{acknowledgement}

This work was supported in part by a grant under Indian Space Research Organization (ISRO), by the grants under Ramanujan Fellowship, Early Career Award, and Nano Mission from the Department of Science and Technology (DST), and by a grant from MHRD, MeitY and DST Nano Mission through NNetRA.

\end{acknowledgement}

\begin{suppinfo}

Supporting Information is available on characterization of photodetector D2.

\end{suppinfo}

\bibliography{reference}
\newpage
\begin{figure}[!hbt]
\centering
%\vs{-0.1in}
%\hs{-1in}
\includegraphics[scale=0.5]{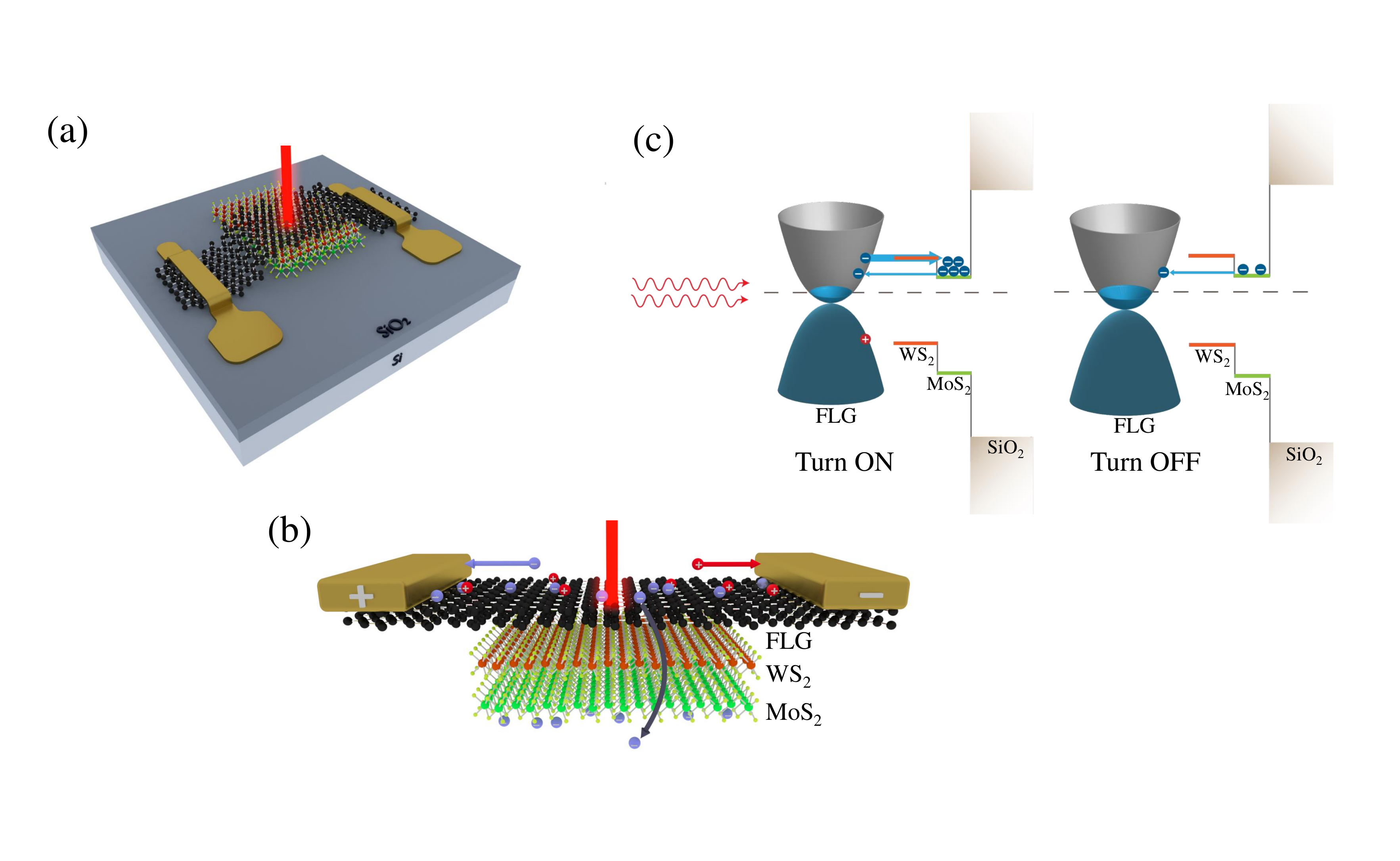}
%\vspace{-1.8in}
\caption{\textbf{Proposed photodetector and principle of operation.} (a) Schematic view of the proposed vertical heterojunction device structure. (b) Schematic view of the cross section of the proposed device showing inter-layer transfer of electrons from few layer graphene (FLG) to MoS$_2$ quantum well under illumination, causing photogating effect. (c) Band diagram along the vertical direction depicting electron exchange between FLG and MoS$_2$ when illumination is turned ON (left panel), creating a steady state population in MoS$_2$ causing photogating. When light is turned OFF (right panel), the stored carriers are pushed back to graphene through thin WS$_2$ tunnel barrier due to built-in field, causing fast turn OFF.}\label{fig:fig1}
\end{figure}
\newpage
\begin{figure}[!hbt]
\centering
%\vs{-0.1in}
%\hs{-1in}
\includegraphics[scale=0.5]{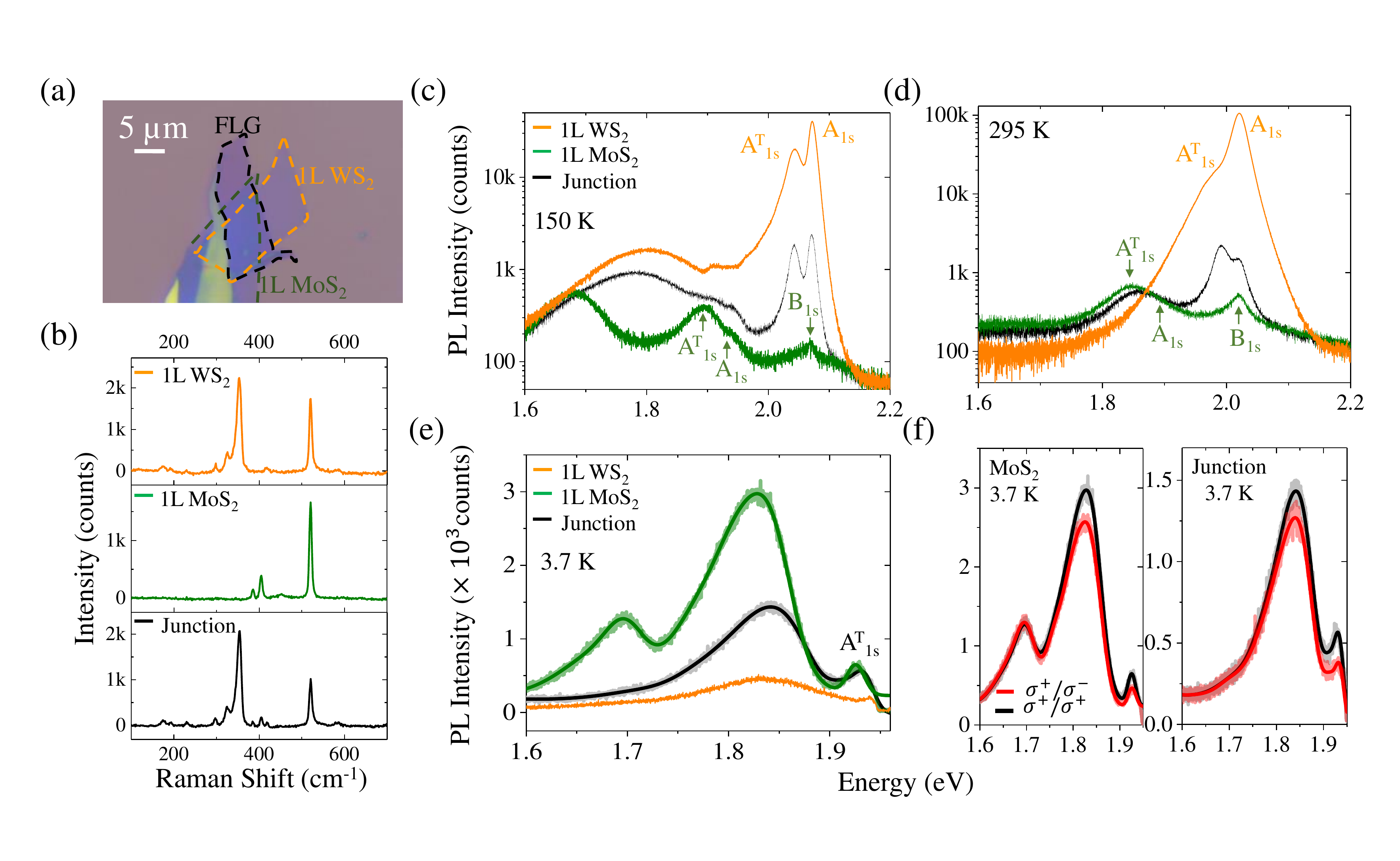}
%\vspace{-1.8in}
\caption{\textbf{Evidence of carrier storage in MoS$_2$ quantum well.} (a) Optical image of a FLG/1L-WS$_2$/1L-MoS$_2$ vertical stack, with isolated portions as controls. Perimeter of each layer is shown by dashed lines. (b) Raman spectra of isolated monolayer regions along with the junction (FLG/1L WS$_2$/1L MoS$_2$) with 532 nm laser excitation. (c-d) PL intensity (in log scale) of the three regions with 532 nm laser excitation at (c) 150 K and (d) 295 K. The individual exciton and charged exciton (trion) peaks are marked. While WS$_2$ peaks show strong PL quenching at the junction, the MoS$_2$ exciton and trion maintain similar luminescence as in the isolated portion, suggesting efficient carrier confinement in MoS$_2$ quantum well. (e) PL intensity of the three different regions with 633 nm laser excitation at 3.7 K. WS$_2$ PL is non-existent due to sub-bandgap excitation. MoS$_2$ trion peak intensity remains unaltered between isolated portion and junction. Strong and broad luminescence below trion peak of MoS$_2$ indicates existence of sub-bandgap defect states. The exciton peak is truncated by the edge filter. (f) Circular polarization resolved PL on isolated MoS$_2$ (left panel) and on junction (right panel) using 633 nm excitation, at 3.7 K. The black line indicates $\sigma^+$ excitation and $\sigma^+$ detection, while the red line indicates $\sigma^+$ excitation and $\sigma^-$ detection. Trion valley polarization contrast is found to $\sim$ 25\% in both cases.
 }\label{fig:fig2}
\end{figure}
\newpage
\begin{figure}[!hbt]
\centering
%\vs{-0.1in}
%\hs{-1in}
\includegraphics[scale=0.5]{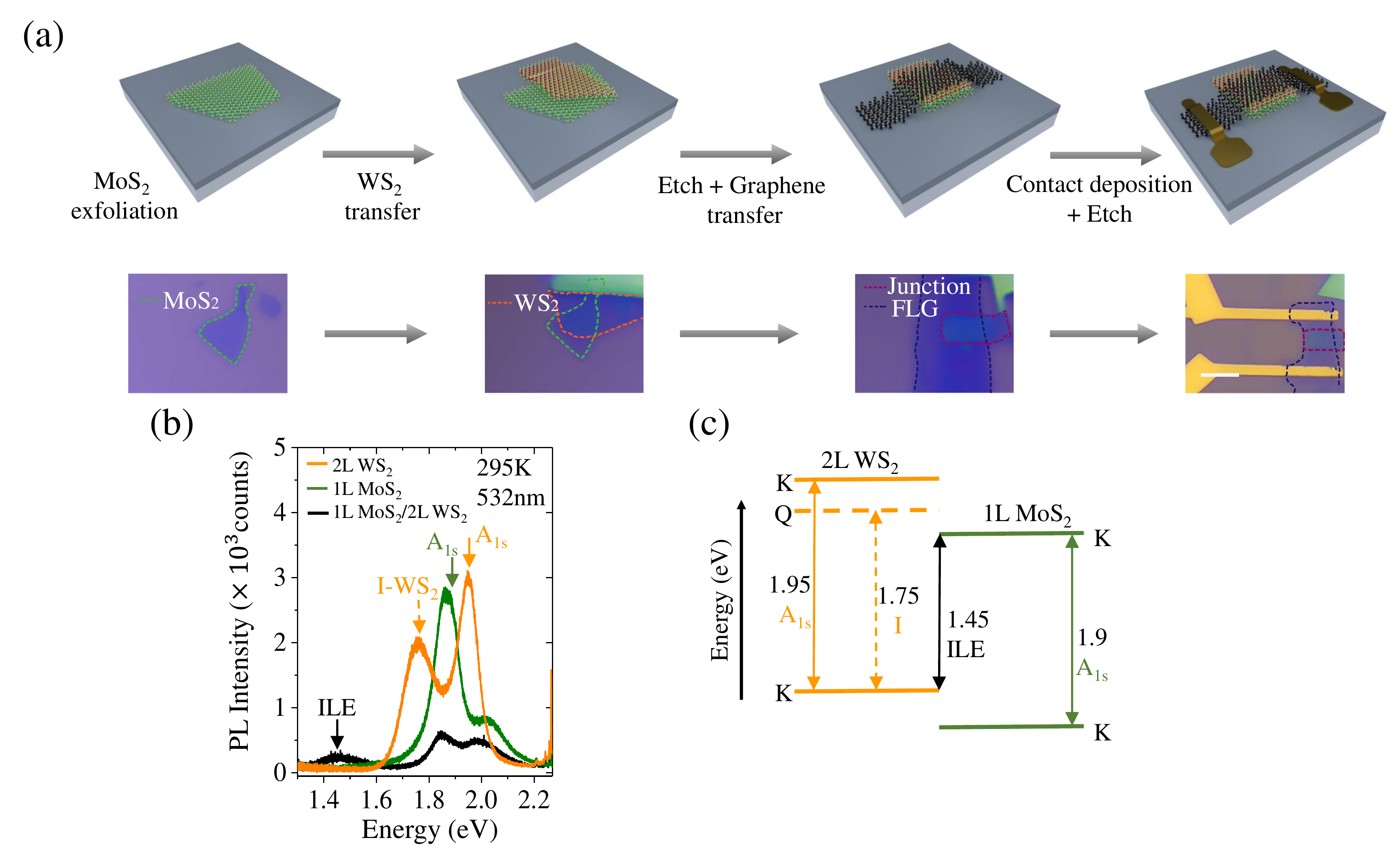}
%\vspace{-1.8in}
\caption{\textbf{FLG/2L-WS$_2$/1L-MoS$_2$ photodetector fabrication.} (a) Process flow for FLG/2L-WS$_2$/1L-MoS$_2$ photodetector fabrication with corresponding optical images at different stages in the bottom row. Scale bar is 5 $\mu$m. (b) PL spectra of 2L-WS$_2$, 1L-MoS$_2$ and 2L-WS$_2$/1L-MoS$_2$ junction. I-WS$_2$ and ILE indicate the indirect peak of WS$_2$ and the inter-layer peak between WS$_2$ and MoS$_2$, respectively. (c) The band alignment of 2L-WS$_2$/1L-MoS$_2$ as obtained from the PL peaks.}\label{fig:fig3}
\end{figure}
\newpage
\begin{figure}[!hbt]
\centering
%\vs{-0.1in}
%\hs{-1in}
\includegraphics[scale=0.5]{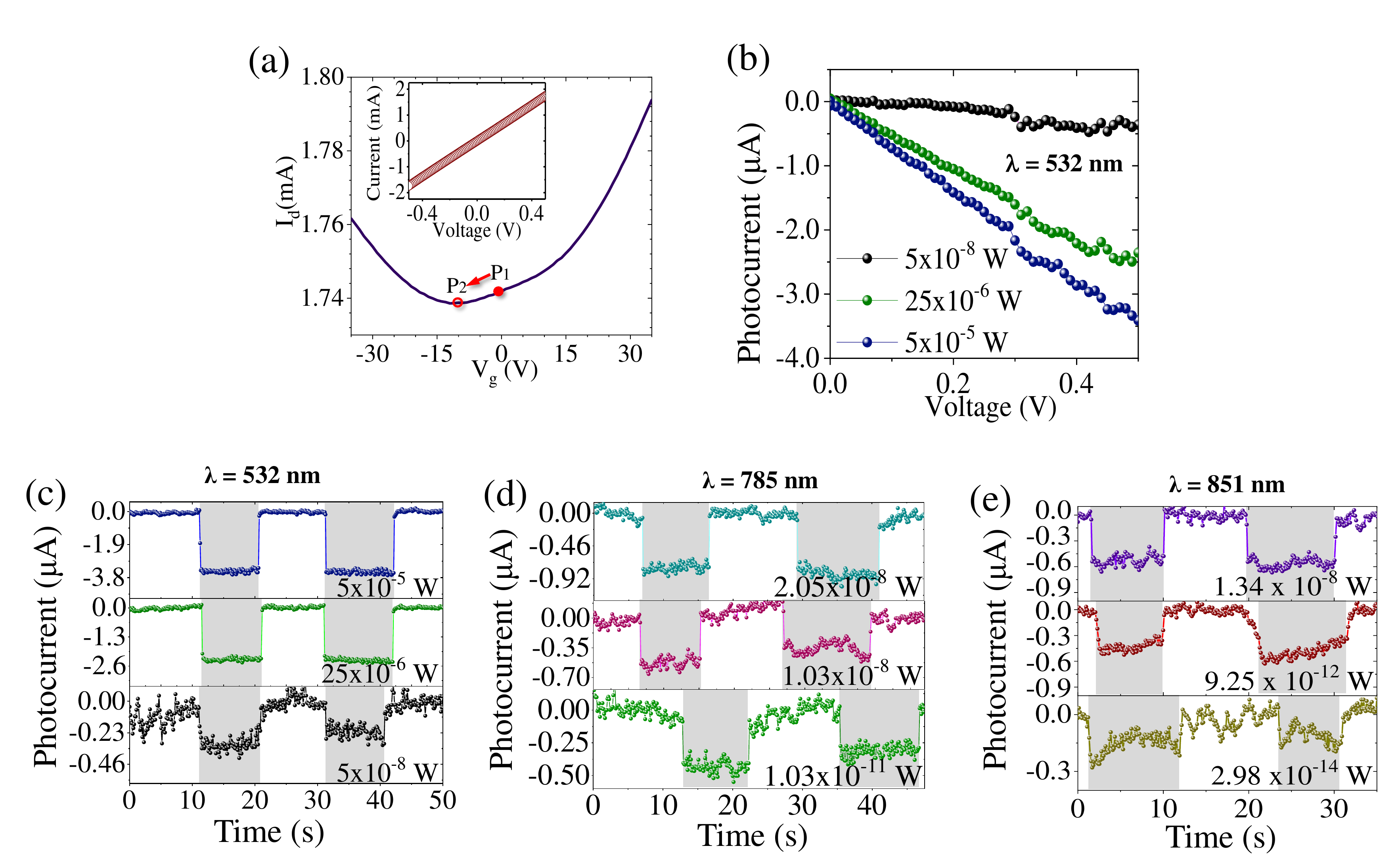}
%\vspace{-1.8in}
\caption{\textbf{Characteristics of FLG/2L-WS$_2$/1L-MoS$_2$ photodetector.} (a) I$_d$-V$_g$ plot for FLG under dark condition with $V_d=0.5$ V. Point $P_1$ (red, solid circle) shows device operating point under dark condition at zero $V_g$ which shifts to point $P_2$ (red, open circle) under illumination. Inset: Drain voltage dependant dark current under zero $V_g$. (b) Photocurrent versus source-drain bias voltage with 532 nm excitation, at three different incident optical powers. (c-e) Transient photoresponse of the proposed graphene photodetector with (c) 532 nm, (d) 785nm and (e) 851 nm excitations at varying optical powers.}\label{fig:fig4}
\end{figure}
\newpage
\begin{figure}[!hbt]
\centering
%\vs{-0.1in}
%\hs{-1in}
\includegraphics[scale=0.5]{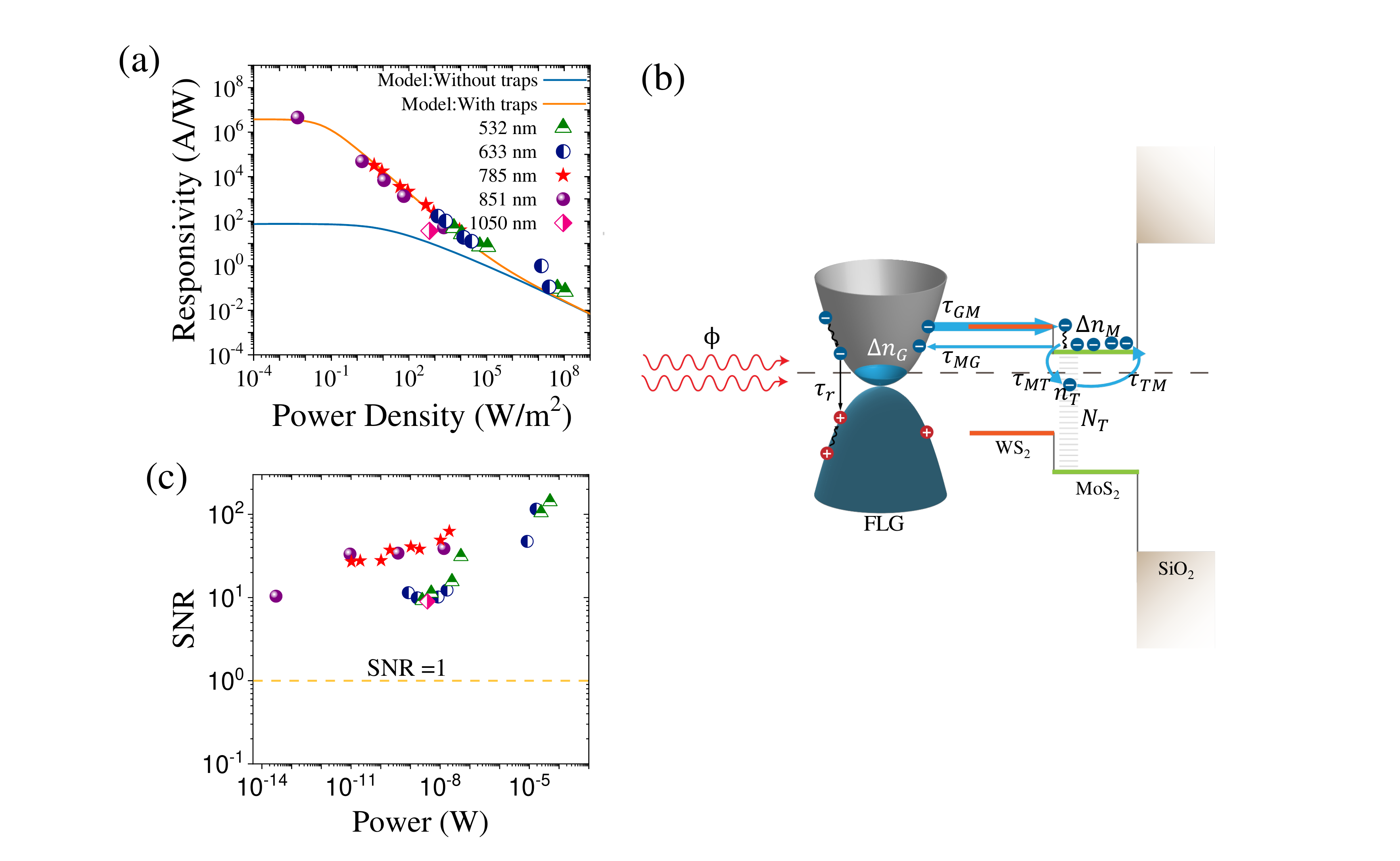}
%\vspace{-1.8in}
\caption{\textbf{Performance of FLG/2L-WS$_2$/1L-MoS$_2$ photodetector.} (a) Measured (symbols) responsivity versus incident optical power density plot with varying wavelengths. At the minimum power density, the extracted responsivity is $4.4\times10^6$ A/W. Solid lines in orange and blue represent the model predicted responsivity with and without trap states in MoS$_2$, respectively. (b) Schematic representation showing different processes captured in the model. (c) Signal-to-noise ratio ($S\!N\!R$) of the detector as a function of the actual power falling on the junction. The dashed line indicates the detection limit of the photodetector with $S\!N\!R = 1$.}\label{fig:fig5}
\end{figure}
%\begin{figure*}[!hbt]
%\centering
%\vs{-1in}
%%\hs{-1in}
%\includegraphics[scale=0.85]{figs/SI.pdf}
%%\vspace{-1.8in}
%%\caption{\textbf{Performance of FLG/2L-WS$_2$/1L-MoS$_2$ photodetector.} (a) Measured (symbols) responsivity versus incident optical power density plot with varying wavelengths. At the minimum power density, the extracted responsivity is $4.4\times10^6$ A/W. Solid lines in orange and blue represent the model predicted responsivity with and without trap states in MoS$_2$, respectively. (b) Schematic representation showing different processes captured in the model. (c) Extracted noise equivalent power ($NEP$) as a function of incident  optical power density. At the lowest power density used, we obtain $NEP$ of 4 fW/$\sqrt{Hz}$.}\label{fig:SI}
%\end{figure*}
\includepdf[pages={1-3}]{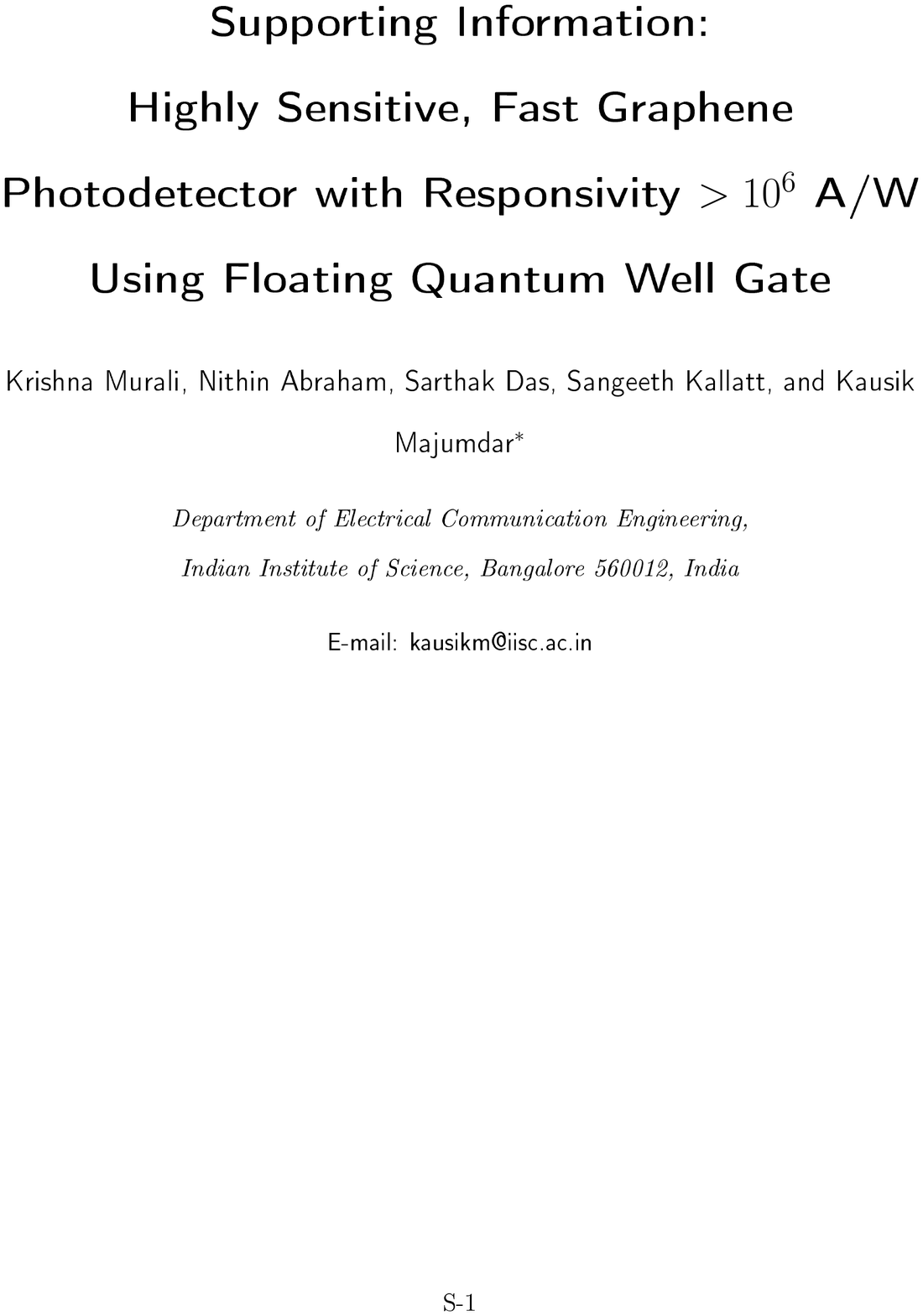}
\end{document}